\documentstyle[12pt]{article}
\begin{document}

\begin{titlepage}

\vspace*{-2cm}


\vspace{.5cm}

\begin{centering}

{\huge Higher order cohomological restrictions on 
anomalies and counterterms} 

\vspace{1cm}

{\large G. Barnich$^*$}

\vspace{1cm}

Freie Universit\"at Berlin, Fachbereich Physik, Institut f\"ur
Theoretische Physik, Arnimallee 14, D-14195 Berlin.

\vspace{1cm}

\begin{abstract}
Using a regularization with the properties of dimensional
regularization, higher order local consistency conditions on
one loop anomalies and 
divergent counterterms are given. They are derived
without any a priori assumption on the form of the BRST cohomology and
can be summarized by the statements that (i) the antibracket 
involving the first order divergent counterterms, respectively the
first order anomaly, with any BRST cocycle is BRST exact, 
(ii) the first order divergent counterterms can be completed into a
local deformation of the solution of the master equation and (iii) the
first order anomaly can be deformed into a local cocycle of the
deformed solution.   
\end{abstract}

\vspace{5cm}

\end{centering}

{\footnotesize \hspace{-0.6cm}($^*$)Alexander von Humboldt fellow. 
Starting November 1st 1997: Charg\'e de Recherches du Fonds National
Belge de la Recherche Scientifique at Universit\'e Libre de Bruxelles,
Facult\'e des Sciences, Campus Plaine C.P. 231, B-1050 Bruxelles,
Belgium.}

\end{titlepage}

\pagebreak
\def\qed{\hbox{${\vcenter{\vbox{                         
   \hrule height 0.4pt\hbox{\vrule width 0.4pt height 6pt
   \kern5pt\vrule width 0.4pt}\hrule height 0.4pt}}}$}}
\newtheorem{theorem}{Theorem}
\newtheorem{lemma}{Lemma}
\newtheorem{definition}{Definition}
\newtheorem{corollary}{Corollary}
\newcommand{\proof}[1]{{\bf Proof.} #1~$\qed$}
\renewcommand{\theequation}{\thesection.\arabic{equation}}
\renewcommand{\thetheorem}{\thesection.\arabic{theorem}}
\renewcommand{\thelemma}{\thesection.\arabic{lemma}}
\renewcommand{\thecorollary}{\thesection.\arabic{corollary}}
\renewcommand{\thedefinition}{\thesection.\arabic{definition}}
\def\be{\begin{eqnarray}}
\def\ee{\end{eqnarray}}
\def\ben{\begin{eqnarray*}}
\def\een{\end{eqnarray*}}


Cohomological techniques in renormalization theory have attracted a
lot of interest since their introduction in the pioneering work
by Becchi, Rouet and Stora \cite{BRS} in the context of Yang-Mills
theories, because they allow to address
general problems of perturbative quantum field theories, like the form
of anomalies or of divergent counterterms by purely
algebraic means (see \cite{PiSo} for a recent review). 
The BRST construction and the formulation of Zinn-Justin
thereof in terms of the generating functional for vertex functions
\cite{Zin} has then been generalized to theories with general
local symmetries by Batalin and Vilkovisky \cite{BaVi}. 

In this more general setting, one loop 
anomalies are constrained by the
cohomology groups in ghost number $1$ of the BRST differential
generated by a solution to the master equation
\cite{HLW,TVV}. This represents the generalization of the Wess-Zumino
consistency condition \cite{WeZu}
for the case of the gauge anomaly. The cohomological 
restrictions on the one loop divergent counterterms, 
involving the ghost number $0$
group, have been discussed in
\cite{Ans,GoWe}, where it is stressed that these techniques are
valid in the power counting non renormalizable case or for the case of
higher dimensional operators (see also
\cite{KlZu}) and hence apply to effective field theories 
\cite{Wei}.

These works only needed the first order consistency
conditions following from the quantum action principle mainly for the
following reason. One can
show \cite{BBH2} that for semi-simple Yang-Mills theory or gravity,
the cohomology groups in ghost number $0$ and $1$
can be described independently of
the antifields, so that all higher order constraints to be found below
turn out to be trivial.
This is however not the case in general, where 
the form of the BRST cohomology groups in ghost number
$0$ and $1$ can be more involved. In this case, 
higher order considerations
will be relevant. 

The purpose of this letter
is to give a purely cohomological description in the space of local
functionals of the higher order
restrictions on the one loop anomalies and counterterms
in terms of deformation theory. To derive these
conditions, we assume that there is a regularization scheme with the
properties of dimensional regularization as used in \cite{Ton},
although we expect the cohomological restrictions to be independent of
the regularization method. 
Notational conventions for the Batalin-Vilkovisky formalism 
will be those of the reviews \cite{HeTe,GPS}. The analysis applies to 
local and rigid symmetries if we understand the master equation to be
the generalized master equation discussed in \cite{BHW}. In this
letter, only the cocyle condition is considered. More details including
a discussion on the coboundary conditions will
appear elsewhere \cite{Bar}.

A different, but related issue is the problem of cohomological restrictions
on anomalies and counterterms at higher loops. Such considerations 
have appeared 
in the recent literature \cite{Whi,Ton,DPT} from various points
of view, and in particular, the form of the 
consistency conditions at higher loops 
has been discussed in \cite{PaTr} in the context
of non local regularization. These issues will be adressed in the
set-up of this letter in
\cite{Bar}.

\section{Regularization}
\setcounter{equation}{0}
We will assume that there is a regularization with the properties of
dimensional regularization as explained in reference \cite{Ton}, i.e.,

\begin{itemize}
\item the regularized action $S_\tau=\Sigma_{n=0}\tau^n S_n$ 
is a polynomial or a power series in
$\tau$, the $\tau$ independent part corresponding to the starting
point action $S_0=S$,

\item if the renormalization has been carried out up to $n-1$ loops, 
the divergences of the effective action at $n$ loops 
are poles in $\tau$ up to the order $n$ with residues that are local
functionals, 
and 

\item the regularized quantum action principle holds \cite{BrMa}.
\end{itemize}

Let $\tilde
S=S_\tau+\rho^*\theta_\tau$, with
$\theta_\tau=\frac{1}{2\tau}(S_\tau,S_\tau)$, so that
$\theta_0=(S,S_1)$, and
$\rho^*$ a global source in ghost number $-1$. On the classical level,
we have, using $(\rho^*)^2=0$,
\be
\frac{1}{2}(\tilde S,\tilde S)=\tau\frac{\partial \tilde S}
{\partial \rho^*},\label{cl}\\
(\tilde S,\frac{\partial \tilde S}{\partial \rho^*})=0,\label{cl1}
\ee
which translates, according to the quantum action principle, 
into the corresponding equations for the regularized generating functional
for 1PI vertex functions:
\be
\frac{1}{2}(\tilde \Gamma,\tilde \Gamma)=\tau\frac{\partial \tilde
  \Gamma}
{\partial \rho^*},\label{actp}\\
(\tilde\Gamma,\frac{\partial \tilde
  \Gamma}
{\partial \rho^*})=0.
\ee

Using $(\rho^*)^2=0$, these equations reduce to 
\be
\frac{1}{2}(\Gamma,\Gamma)=\tau\frac{\partial \tilde
  \Gamma}
{\partial \rho^*},\label{3}\\
(\Gamma,\frac{\partial \tilde
  \Gamma}
{\partial \rho^*})=0.\label{2}
\ee
At one loop, we get 
\be
(S_\tau,\Gamma^{(1)})=\tau\theta^{(1)},\label{g}\\
(S_\tau,\theta^{(1)})+(\Gamma^{(1)},\theta_\tau)=0,\label{l}
\ee
where $\Gamma^{(1)}$ and $\theta^{(1)}$ are respectively the one loop
contributions of $\Gamma$ and $\partial \tilde
  \Gamma/{\partial \rho^*}$. By assumption, we have
\be
\Gamma^{(1)}=\sum_{n=-1}\tau^n\Gamma^{(1)n},\\ 
\theta^{(1)}=\sum_{n=-1}\tau^n\theta^{(1)n},
\ee
where $\Gamma^{(1)-1}$
and $\theta^{(1)-1}$ are local functionals.

\section{Lowest order cohomological restrictions}
\setcounter{equation}{0}

At ${1}/{\tau}$, equations (\ref{g}) and (\ref{l}) give
\be
(S,\Gamma^{(1)-1})=0\label{3.4},\\
(S,\theta^{(1)-1})+(\Gamma^{(1)-1},\theta_0)=0\label{3.3}.
\ee
Using 
$\theta_0=(S,S_1)$ and equation (\ref{3.4}), equation (\ref{3.3}) 
reduces to 
\be
(S,\theta^{(1)-1}-(S_1,\Gamma^{(1)-1}))=0.
\ee
In addition, we get, from the term independent of $\tau$ in 
equation (\ref{g}),
\be
(S,\Gamma^{(1)0})=\theta^{(1)-1}-(S_1,\Gamma^{(1)-1})
\label{3.6}.
\ee
The term linear in $\tau$ gives
\be
(S,\Gamma^{(1)1})=\theta^{(1)0}-(S_1,\Gamma^{(1)0})
-(S_2,\Gamma^{(1)-1}).
\ee
The one loop renormalized effective action is 
$\Gamma^1_R=
S+\hbar \Gamma^{(1)0}+O(\hbar^2,\tau)$,
where the notation $O(\hbar^2,\tau)$ means that those terms which are
not of order at least two in $\hbar$ are of order at least one in
$\tau$ and vanish when the regularization is removed
($\tau\longrightarrow 0$), so that 
\be
{1\over 2}(\Gamma^1_R,\Gamma^1_R)=\hbar A_1+O(\hbar^2,\tau),
\ee
with, using equation (\ref{3.6}),
\be
A_1=\theta^{(1)-1}-(S_1,\Gamma^{(1)-1}),\label{a}
\ee 
and the consistency conditions for local functionals
\be
(S,\Gamma^{(1)-1})=0\Longrightarrow \Gamma^{(1)-1}
=c_1^iC_i+(S,\Xi_1),\\
(S,A_1)=0\Longrightarrow A_1=a_1^iA_i+(S,\Sigma_1),
\ee
where $C_i$ and $A_i$ are respectively a basis of representatives for
$H^{0}(s)$ and $H^{1}(s)$. 

\section{First order cohomological restrictions}
\setcounter{equation}{0}
\setcounter{theorem}{0}

Let $D$ be a BRST cocycle in any ghost number $g$ and consider 
$S^j=S+jD$, where the source $j$ is of ghost number $-g$. The
regularized action is $S^j_\tau=S_\tau+jD_\tau$, with $D$ a polynomial
in $\tau$ starting with $D$. If $\tilde S^j=S^j_\tau+\rho^*\theta^j_\tau$,
where $\theta^j_\tau={1\over2\tau}(S_\tau,S_\tau)+{1\over\tau}
(jD_\tau,S_\tau)$, we have
\be
{1\over 2}(\tilde S^j,\tilde S^j)=\tau 
{\partial \tilde S^j\over\partial\rho^*}+O(j^2),
\ee
and the corresponding equation for the regularized generating
functional $\tilde \Gamma_j$
\be
{1\over 2}(\tilde \Gamma_j,\tilde \Gamma_j)=\tau 
{\partial \tilde \Gamma_j\over\partial\rho^*}+O(j^2).
\ee
At one loop, we get for the term independent of $\rho^*$,
\be
(\Gamma^{(1)}_j,S^j_\tau)=\tau\theta^{(1)}_j+O(j^2).
\ee
The term linear in $j$ of order ${1\over\tau}$ gives
\be
(D^{(1)-1},S)+(D,\Gamma^{(1)-1})=0,\label{heur}
\ee
with $D^{(1)-1}=(\partial\Gamma^{(1)-1}_j/\partial j)|_{j=0}$.
This gives our first theorem.
\begin{theorem}
The antibracket of the divergent one loop part $\Gamma^{(1)-1}$,
which is BRST closed and local, 
with any local BRST cocycle is BRST exact in the
space of local functionals.
\end{theorem}

The theorem
can be reformulated by saying that 
the antibracket map (induced in the local BRST cohomology groups by the
antibracket, see ref.~\cite{BaHe}) 
\be
([\Gamma^{(1)-1}],[D])=[0]
\ee
for all $[D]\in H^g(s)$.
This equation represents a cohomological
restriction on the coefficients $c^i_1$ that can appear~; it can be
calculated classically from the knowledge of $H^0(s)$ and the
antibracket map  from $H^0(s)\times H^g(s)$ to
$H^{g+1}(s)$. 
According to the previous section, the theorem holds in particular
when $D=\Gamma^{(1)-1}$ or $D=A_1$. 

In the same way, the consistency condition is
\be
(\Gamma_j,{\partial \tilde \Gamma_j\over\partial\rho^*})+O(j^2)=0,
\ee
and gives at one loop,
\be
(\Gamma^{(1)}_j,\theta^j_\tau)+(S^j_\tau,\theta^{(1)}_j)+O(j^2)=0.
\ee
The term linear in $j$ of order ${1\over\tau}$ gives
\be
(D^{(1)-1},\theta_0)-\left(\left.{\partial {\theta^j}_0\over\partial j}
\right|_{j=0},
\Gamma^{(1)-1}\right)\nonumber\\+(D,\theta^{(1)-1})
-\left(\left.{\partial \theta_j^{(1)-1}\over\partial j}\right|_{j=0},S
\right)=0.
\ee
Using $\theta_0=(S,S_1)$, ${\partial \theta^j_0/\partial
  j}|_{j=0}=(D_1,S)+(D,S_1)$,  equations (\ref{3.4}), (\ref{a})
and (\ref{heur}),
we get
\be
(D,A_1)-\left(\left.{\partial \theta_j^{(1)-1}\over\partial
    j}\right|_{j=0}-(D_1,\Gamma^{(1)-1})-(D^{(1)-1},S_1),S\right)=0.
\label{3.9}
\ee
This gives our second result.
\begin{theorem}
The antibracket of the BRST closed first order anomaly $A_1$
with any local BRST cocycle is BRST exact in the
space of local functionals.
\end{theorem}

The theorem
can again be reformulated by saying that 
the antibracket map 
\be
([A_1],[D])=[0]
\ee
for all $[D]\in H^g(s)$~; it represents a classical cohomological
restriction on the coefficients $a^i_1$ that can appear. 

\section{Higher orders}
\setcounter{equation}{0}

Let $B^0=S$ and $B^1=\Gamma^{(1)-1}$. We have the following theorem.
\begin{theorem}
The first order counterterms can be completed into a local deformation
of $S$, i.e., there exist local functionals $B^n$ such that 
\be 
\frac{1}{2}(S^{j^\infty},S^{j^\infty})=0,\\
S^{j^\infty}=S+\sum_{n=1}j^nB^n.
\ee
\end{theorem}
The higher order cohomological restrictions of such an equation in
terms of Lie-Massey brackets is briefly discussed in \cite{BaHe}. More
details will be given in 
\cite{Bar}. 
 
\proof{The theorem is true for $j^0,j^1$ and $j^2$, if we take
$D=\Gamma^{(1)-1}=B^1$ in (\ref{heur}) and $B^2=1/2
(\partial\Gamma^{(1)-1}_j/\partial j)|_{j=0}$. 
Suppose the theorem true at order $j^k$ i.e., we have  
\ben
\frac{1}{2}(S^{j^k},S^{j^k})=O(j^{k+1}),
\\S^{j^k}=S+\sum^k_{n=1}j^nB^n.
\een
and 
\ben
B^n={1\over n}(\partial^{n-1}\Gamma^{(1)-1}_{j^{n-1}}/\partial j^{n-1})|_{j=0}.
\een
At the regularized level, consider the action 
\ben
S_\tau^{j^k}=S_\tau+\sum^k_{n=1}j^nB_\tau^n
\een
and $\tilde
S^{j^k}=S_\tau^{j^k}+\rho^*\theta_{\tau}^{j^k}$, with
$\theta_{\tau}^{j^k}=\frac{1}{2\tau}
(S_\tau^{j^k},S_\tau^{j^k})+O(j^{k+1})$, so that
\ben
\frac{1}{2}(\tilde S^{j^k},\tilde S^{j^k})=
\tau\frac{\partial\tilde S^{j^k}}{\partial
\rho^*}+O(j^{k+1}). 
\een
The corresponding equation for 
$\tilde \Gamma_{j^k}$ based on the action $\tilde S^{j^k}$ is 
\ben
\frac{1}{2}(\tilde \Gamma_{j^k},\tilde \Gamma_{j^k})=
\tau\frac{\partial\tilde \Gamma_{j^k}}{\partial
\rho^*}+O(j^{k+1}). 
\een
At one loop, we get, for the part independent of
$\rho^*$,  
\ben
(S_\tau^{j^k},\Gamma^{(1)}_{j^k})=
\tau\theta^{(1)}_{j^k}+O(j^{k+1}).
\een
At order $j^k$, this equation gives 
\ben
\left (S_\tau,\left.{\partial^k
\Gamma^{(1)}_{j^k}\over\partial j^k}\right|_{j=0}\right )
+\left (B_\tau^1,\left.{\partial^{k-1}
\Gamma^{(1)}_{j^k}\over\partial j^{k-1}}\right|_{j=0}\right )
+\dots\nonumber\\+\left
(B_\tau^k,\left.\Gamma^{(1)}_{j^k}\right|_{j=0}\right )=
\tau\left.{\partial^k
\theta^{(1)}_{j^k}\over\partial j^k}\right|_{j=0}.
\een
At order $1/\tau$, we get, using 
\ben
\left.{\partial^{n-1}
\Gamma^{(1)-1}_{j^k}\over\partial j^{n-1}}\right|_{j=0}=
\left.{\partial^{n-1}
\Gamma^{(1)-1}_{j^{n-1}}\over\partial j^{n-1}}\right|_{j=0}=nB^n,
\een
for $n=1,\dots,k-1$ and defining $\left.{\partial^k
\Gamma^{(1)-1}_{j^k}\over\partial j^k}\right|_{j=0}=(k+1)B^{k+1}$, 
the relation
\ben
(S,(k+1)B^{k+1})+(B^1,kB^k)+\dots+(B^{k},B^1)=0,
\een
or equivalently
\be
0=\sum_{m=0}^k(B^m,(k+1-m)B^{k+1-m})=
{(k+1)\over 2}\sum_{m=0}^{k+1}(B^m,B^{k+1-m}),\label{id}
\ee
which proves the theorem.
} 

\vspace{1cm}

Let $E^0=A_1=\theta^{(1)-1}-(B^1,S_1)$.
\begin{theorem}
The lowest order contribution to the anomaly $E^0$ can be extended to
a local cocycle of the deformed solution of the master equation
$S^{j^\infty}$, i.e., there exist local functionals $E^m$ such that 
\be
(S^{j^\infty},E^{j^\infty})=0,\\
E^{j^\infty}=\sum_{m=0}j^mE^m.
\ee
\end{theorem}

\proof{The theorem holds for $j^0$ and $j^1$ by taking in (\ref{3.9})
$D=B^1$, and defining 
\ben
E^1=\left.{\partial \theta_j^{(1)-1}\over\partial
    j}\right|_{j=0}-(D_1,\Gamma^{(1)-1})-(D^{(1)-1},S_1)\\=
\left.{\partial \theta_j^{(1)-1}\over\partial
    j}\right|_{j=0}-(B_1,B^1_1)-(2B^2,S_1).
\een
Let us define
\be
E^m=\left.{\partial^{m} \theta_{j^{m}}^{(1)-1}\over\partial
   j^m}\right|_{j=0}-\sum_{n=0}^{m}((n+1)B^{n+1},B^{m-n}_1).\label{def}
\ee
The consistency condition is 
\ben
(\Gamma_{j^k},{\partial \tilde\Gamma_{j^k}\over\partial \rho^*})=O(j^{k+1}).
\een
At one loop, we have,
\ben
(\Gamma^{(1)}_{j^k},\theta_\tau^{j^k})+
(S_\tau^{j^k},\theta_{j^k}^{(1)})=O(j^{k+1}).
\een
The term of order $j^k$ of this equation gives 
\ben
\sum_{m=0}^{k}
\left[\left(\left.{\partial^m \Gamma^{(1)}_{j^k}\over\partial j^m}
\right|_{j=0},\left.{\partial^{k-m} \theta_\tau^{j^k}\over\partial j^{k-m}}
\right|_{j=0}\right)
+\left(B^m_\tau,\left.{\partial^{k-m} \theta_{j^k}^{(1)}\over\partial j^{k-m}}
\right|_{j=0}\right)\right]=0.
\een
At order $1/\tau$, we get 
\be
\sum_{m=0}^{k}\Bigg[\left((m+1)B^{m+1},
\sum_{l=0}^{k-m}(B^l,B^{k-m-l}_1)\right)
\nonumber\\
+\left (B^m,\left.{\partial^{k-m} \theta_{j^k}^{(1)-1}\over\partial j^{k-m}}
\right|_{j=0}\right )\Bigg]=0.\label{bas}
\ee
Using the Jacobi identity, the first term is given by 
\ben
\sum_{m=0}^{k}\sum_{l=0}^{k-m}\left[\left(\left((m+1)B^{m+1},
B^{k-m-l}\right),B^l_1\right)
-\left(B^l,\left((m+1)B^{m+1},B^{k-m-l}_1\right)\right)\right].
\een
Changing the sum $\sum_{m=0}^{k}\sum_{l=0}^{k-m}$ to the equivalent 
sum $\sum_{l=0}^{k}\sum_{m=0}^{k-l}$, the first term of this equation
vanishes on account of (\ref{id}), while the second term, using the
definition (\ref{def}), combines with the second term of (\ref{bas}) to
give
\ben
\sum_{m=0}^{k}(B^m,E^{k-m})=0,
\een
which proves the theorem.
}

The investigation in this letters is a 
first step in order to analyze
the cohomological restrictions on anomalies and counterterms at higher
orders in $\hbar$. To see this, we note that if we put
$j=(-\hbar/\tau)$, the action $S^{(-\hbar/\tau)^\infty}$ satisfies the
(deformed) master equation
$1/2(S^{(-\hbar/\tau)^\infty},S^{(-\hbar/\tau)^\infty})=0$, while the
correponding effective action is finite at order $\hbar$. Its
divergences at order $\hbar^2$ are poles up to order $2$ in $\tau$
with residues that are local functionals. A systematic analysis of the
substraction procedure at higher orders in $\hbar$ will be presented
in \cite{Bar}.

\section*{Acknowledgments}

The author wants to thank F.~Brandt, M.~Henneaux, J.~Kottmann,
M.~Kreuzer, J.~Par\'{\i}s and 
M.~Tonin for useful discussions and Prof. H. Kleinert for
hospitality in his group while this work has been completed.

\end{document}